\documentclass[preprint,aps]{revtex4}
\usepackage{graphicx,graphics,color,epsfig}
\begin{document}
\title{Quantum key distribution with asymmetric channel noise}
\author{Xiang-Bin Wang}\email{wang@qci.jst.go.jp}
 \affiliation{Imai Quantum Computation and Information Project, ERATO, 
JST,
Hongo White Building 201, 5-28-3, Hongo, Bunkyo, Tokyo 113-0033, Japan}
\date{\today}

\begin{abstract}We show that, one may take advantage of the asymmetry of
channel noise. With appropriate modifications to the standard protocols, both
the
key rate and the tolerable total channel noise can be increased if
the channel noise is asymmetric.
\end{abstract}
\maketitle \section {Introduction} Unlike classical key distribution, quantum
key distribution (QKD) is built on the fact that measuring an unknown quantum
state will almost surely disturb the state.  The first
published QKD protocol, proposed in by Bennett and Brassard in 1984\cite{BB},
is called BB84. For a history of the subject, one
may refer to, for example, Ref.~\cite{gisin}. Since then, studies on QKD are
extensive.  Strict mathematical proofs for the unconditional security have
been given already\cite{qkd,mayersqkd,others,others2,shorpre}. It
is greatly simplified if one connects this with the quantum
entanglement purification protocol
(EPP)\cite{BDSW,deutsch,qkd,shorpre}.

To all QKD protocols, the first important requirement is the 
unconditional security, i.e., using whatever type of 
attack, eavesdropper (Eve) can never have non-negligible information
to the final key shared by Alice and Bob. A conditional secure
protocol without a testable condition is not so useful, since
it is essentially same to the classical protocol on its security.
 In particular, one cannot
depend the security on certain specific properties of physical channels,
since those specific properties may be changed by Eve at the time the QKD 
protocol is running. For example, based on certain properties of the physical
channel,  Alice and Bob may take some
physical treatment to decrease the error rate of their physical channel. 
Given such a treatment, suppose
the error rate of the improved physical channel is given
by the function of $f(E_0)$, and $E_0$ is the error rate of the 
originally physical channel. Suppose in the QKD protocol Alice and Bob
use the improved physical channel. 
It will be insecure if Alice and Bob 
in their QKD protocol 
only test the errors of the original physical and then use function $f$ to
calculate the error rate for the improved channel and then use the calculated
value as the error rate. This is because, Eve may change the properties of the
original physical channel and the mapping $f$ could be incorrect then.
However, it will be $secure$ if Alice and Bob directly 
test the error rate of the 
improved physical channel and use the tested results as the input of error
correction and privacy amplification. 

Besides the security requirement, we also hope to improve the feasibility
of a protocol in practical use, e.g., the larger tolerable channel error
rate and the key rate. Different from the security issue, 
one can improve the feasibility of a protocol based on the 
properties of physical channel.  Alice and Bob may first
investigate certain properties of their physical channel and then modify
the protocol according to those properties of their physical channel. 
Given the specific
properties of the physical channel, the modified protocol will have certain
advantage, e.g., a lower error rate in their error test.
Note that here they should first be sure that the modified QKD protocol is
unconditionally secure $even$ those assumed properties of physical channel
don't exist.  An $unconditional$ secure 
protocol with certain $conditional$ advantage is allowed in QKD.
That is to say, after taking the modification, we expect certain advantageous
result of the error test. In running the modified protocol,
after the error test step, if Alice and Bob
 find that the result is close
to their expected result, the advantage holds; if Alice and Bob find that
the result is 
quite different from the  expected one, the advantage is seriously
undermined or even
totally lost, but the protocol is still secure. 
In case they find the error test
result  quite different from the expected, their only loss is in the issue
of key rate. Conditional advantage is useful if we have substantial
 probability that the
 error test result in QKD is close to the expected one. 
 Although Eve may in principle do whatever
to change the original physical properties, if Eve wants to hide her presence,
she must not significantly change the error test result of the QKD protocol,
 since 
Alice and Bob will judge that there must be an Eve if their
error test result is too much different from the expected result.
Moreover, the final key rate is only dependent on
the error test results rather than the full properties
of the channel. If the QKD protocol itself is unconditionally secure,
two $different$ channels are $equivalent$ for the QKD purpose
provided that they cause the same error test result.  
This is to say, given an unconditionally
secure protocol, if Eve hides her presence,  Eve's action does not
affect the error test result done by Alice and Bob, i.e., Eve's channel 
is $equivalent$ to the supposed physical channel. The conditional advantage
$always$  holds if Eve hides herself. If Eve does not 
hide herself,
she allows her action to cause the error test result much different from
the expected one. 
 This may  
decrease
the final key rate, or even  destroy the protocol if the error test result is
too far away from the expected one. Even in such a case Eve cannot obtain a nonnegligible 
amount of
information to the final key, given an unconditionally secure protocol.
In this work, in evaluating the feasibility of our protocol, we only 
consider an invisible Eve, i.e., Eve always hides
her presence. Note that no protocol can work as efficiently as
it is expected with a visible Eve. A visible Eve can always destroy 
{\it any} protocol. 

We shall propose protocols with higher  key rate and larger
channel error rate threshold, given an asymmetric physical channel and
invisible Eve.
The key rate and channel error threshold of our protocol are dependent
on the physical channel itself, but the security is independent of
the physical channel, i.e., our protocols are unconditionally secure under
whatever attack outside Alice's and Bob's labs. 
     
Most of the known prepare-and-measure protocols assume the
symmetric channel to estimate the noise threshold for the
protocol. Also, most of the protocols use the symmetrization
method: Alice randomly chooses bases from certain set to
prepare her initial state. All bases in the set is chosen with
equal probability. In such a way, even the noise of the
channel (Eve's channel) is not symmetric, the symmetrization make
the error rate to the key bits be always symmetric. In this work
we show that actually we can let all key bits prepared in a single
basis and we can have advantages in key rate or noise threshold
provided that the channel noise is asymmetric.
\section {Channel error, tested error and key-bits error}
Normally, Alice will
transmit qubits in different basis (e.g., Z basis and X basis) to Bob, 
Bob will also measure
them in different basis. 
The Hadamard transformation 
$H = \frac{1}{\sqrt{2}} \left(\begin{array}{rr}1 & 1 \\
1 & -1 \end{array} \right) $ interchanges the $Z-$basis
$\{|0\rangle,|1\rangle\}$  and $X-$basis
$\{|\pm\rangle=\frac{1}{\sqrt 2}(|0\rangle\pm |1\rangle)\}$.
    We
shall use the term key bits for those raw bits which are used to
distill the final key and the term check bits for those bits whose
values are announced publicly to test the error rates.
Our purpose is to know the bit-flip rate and phase-flip rate
to key-bits. We do it in this way: first test flipping rates
of the check-bits, then deduce the channel flip rate and then
determine
the error rates of key-bits.
As it was shown in Ref\cite{gl}, the flipping rates of qubits
in different bases are in general
different, due to the basis
transformation. Here we give a more detailed study on this issue. 
We first consider the 4-state protocol with CSS code, where only two
basis, $Z-$basis  
and $X-$basis  are involved in operation.
 For such a case of 4-state protocol,
we define asymmetric channel as the channel with its bit flip
error rate being different from its phase flip error rate. 
The check bits will be discarded after the error test. We use the
term Z-bits for those qubits which are prepared and measured in Z
basis, the term $X$-bits
for those bits which are prepared and measured in X basis.
 For clarity we shall regard Alice's action of preparing a 
state in X basis as the joint actions of state preparation
in Z basis followed by a Hadamard transform. We shall also 
regard Bob's measurement in X basis as the joint action of first taking a
Hadamard transform and then taking  the measurement in Z basis.
Therefore we shall 
also call those Z-bits as
$I-$qubits and those $X$-bits as H-bits.
 To let the CSS code work properly, we need to know the
value of average bit-flip rate and average phase-flip rate over 
all key bits. 
We  define three Pauli matrices:
\[
\sigma_x = \left(\begin{array}{rr}0 & 1 \\ 1 & 0 \end{array}
\right), \quad \sigma_y = \left(\begin{array}{rr}0 & -i \\ i & 0
\end{array} \right), \quad \sigma_z = \left(\begin{array}{rr}1 & 0
\\ 0 & -1 \end{array} \right).
\]
The matrix $\sigma_x$ applies a bit flip error but no phase flip
error to a qubit, $\sigma_z$ applies a phase flip error but no bit
flip error, $\sigma_y$ applies both errors. We assume the
$\sigma_x,\sigma_y,\sigma_z$ rates of the physical channel are
$q_{x0},q_{y0},q_{z0}$ respectively. 
Note that the phase flip rate or bit flip
rate of the channel is the summation of $q_{z0},q_{y0}$ or
$q_{x0},q_{y0}$, respectively. Explicitly we have
\begin{eqnarray}
\nonumber p_{x0}=q_{x0}+q_{y0}\\
p_{z0}=q_{z0}+q_{y0}\\
p_{y0}=q_{x0}+q_{z0}
\end{eqnarray}
and $q_{y0}$ is defined as the channel flipping rate to the qubits
prepared in $Y$-basis ($\frac{1}{\sqrt 2}(|0\rangle\pm i|1\rangle)$).
After the error test on check bits, Alice and Bob know the value of
$p_{z0},p_{x0}$ of the channel. Given the channel flipping rates
$q_{x0},q_{y0},q_{z0}$, one can calculate the bit-flip rate and
phase-flip rate for the remained qubits. For those $I-$bits, the
bit-flip rate and phase-flip rate are just $q_{x0}+q_{y0}$ and
$q_{z0}+q_{y0}$ respectively, which are just the channel bit flip
rate and phase flip rate. 
Therefore the channel bit-flip rate $p_{x0}$ is identical
to the tested flip rate of $I-$bits.
The channel phase-flip rate $p_{z0}$ can be determined by testing
the flip rate of those $H-$bits.
 An 
$H-$qubit is a qubit treated in the following order
\\{\it prepared in  $Z$ basis, Hadamard transform, transmitted over the noisy 
channel, Hadamard transform, measurement in $Z$ basis.}
If the channel noise offers a $\sigma_y$  error, the net effect is
\begin{eqnarray}
H \sigma_y H \left(\begin{array}{c}|0\rangle\\|1\rangle\end{array}\right)
=\sigma_y \left(\begin{array}{c}|0\rangle\\|1\rangle\end{array}\right).
\end{eqnarray}
 This shows, the channel $\sigma_y$ error
will also cause a $\sigma_y$ error to an $H-$qubit.
Similarly, due to the fact of
\begin{eqnarray}
H\sigma_z H=\sigma_x\nonumber\\
H\sigma_x H =\sigma_z,
\end{eqnarray}
 a channel $\sigma_x$ flip or a channel
$\sigma_z$ flip will cause a net $\sigma_z$ error or $\sigma_x$ error,
 to an $H-$bit. 
Consequently, a channel phase flip causes a bit flip error to H-bit,
a channel bit-flip causes a phase flip error to H-bit. This is to say,
{\it the measured error of H-bits is just the channel phase flipping rate.} 
Therefore the average 
bit-flip error rate and phase-flip error rate to each types of key bits
will be
\begin{eqnarray}
 p_z^I=p_x^H=p_{z0}\nonumber,\\
p_z^H=p_{x}^I=p_{x0}
\end{eqnarray}
 Here 
$p_x^H,p_x^I$($p_z^H,p_z^I$) are for
the bit flip (phase flip) error of $H$-bits and $I$-bits from
those key-bits respectively.
Suppose the key bits consist of $\eta$ $I$-bits and $1-\eta$ $H$-bits,
the average flip error of the key bits is
\begin{eqnarray}
p_x=\eta p_x^I+(1-\eta)p_z^H=\eta p_{x0}+(1-\eta)p_{z0};\nonumber\\
p_z==\eta p_z^I+(1-\eta)p_x^H=\eta p_{z0}+(1-\eta)p_{x0}.
\end{eqnarray}
\section{ Key rate of QKD protocols with one way communication.} 
We first consider an almost
trivial application of our analysis of asymmetric channel above.
 In the standard BB84 protocol, 
since the preparation 
basis of key bits are symmetrized, the average bit flip error
and phase flip error to those key bits 
are always equal no matter whether the channel
noise itself is symmetric or not. That is to say, when half of
the key-bits are $X$-bits and Half of them are Z-bits, the average
flip  rates over all key bits are always 
\begin{eqnarray}
p_x=p_z=(p_{x0}+p_{z0})/2.
\end{eqnarray}
Therefore the key rate for the standard BB84 protocol
(Shor-Preskill protocol)\cite{shorpre}  with whatever asymmetric
channel is
$1-2H(\frac{p_{x0}+p_{z0}}{2})$\cite{shorpre}, where
$H(t)=-(t\log_2t+(1-t)\log_2(1-t))$. (Note that in the 4-state protocol,
asymmetric channel is simply defined by $p_{x0}\not= p_{z0}$.)
However, if all key bits had
been prepared and measured in $Z$ basis, then the bit flip and
phase flip rates to key bits would be equal to those flipping
values of the channel itself. In such a case
 the  key rate is 
\begin{eqnarray}
R=1-H(p_{x0})-H(p_{z0}).\label{unrate}
\end{eqnarray} 
Obviously, this is, except
 for the special case of $p_{x0}=p_{z0}$, always
 larger than the key rate in standard BB84 protocol with CSS code
(Shor-Preskill protocol), where the key-bits are prepared in 
Z-basis and $X$-basis with equal
 probability. For a higher key rate, one should $always$ use the above modified BB84 protocol with
all key bits prepared and measured in a single basis, Z-basis.

In fact, this type of 4-satte QKD protocol with one single 
basis for key bits had been proposed already
in the past for different purposes, see e.g., ref\cite{onep}. 

Now we  consider the case of 6-state protocol.
In the standard protocol\cite{6state}, the key bits are equally distributed
over all 3 different bases. In distilling the final key, each type of
flipping error used is the averaged value over 3 bases, i.e.,
\begin{eqnarray}\label{average}
 \bar q_x=\frac{q_{x0}+2q_{z0}}{3},
\bar q_z=\frac{q_{x0}+q_{z0}+q_{y0}}{3},
 \bar q_y=\frac{q_{x0}+2q_{y0}}{3}. 
\end{eqnarray} 
The key rate is 
\begin{eqnarray}
r=1-H(\bar q)\nonumber\\
 H(\bar q)=-\bar q_x \log_2 \bar q_x - 
\bar q_y \log_2 \bar q_y 
- \bar q_z \log_2 \bar q_z - q_{I0} \log_2 q_{I0}.
\label{6mix}
\end{eqnarray}
This is the key rate for standard 6-state protocol
where we mix all qubits in diffent bases together.
However, such a mixing is unnecessary. We can choose to simply distill
3 batches of final keys from $Z-$bits, $X-$bits and $Y-$bits separately.
If we do it in such a way, the key rate will be increased to
\begin{eqnarray}
r'=1-q_{x0}\log_2 q_{x0}-q_{y0}\log_2 q_{y0}-q_{z0}\log_2 q_{z0}-
q_{I0}\log_2 q_{I0}.\label{6sep}
\end{eqnarray} 
Obviously, $r'$ is $never$ less than $r$ since the mixing 
operation never decreases the entropy. The only case where
$r'=r$  is $q_{x0}=q_{y0}=q_{z0}$. Therefore for a higher key rate,
we propose to $always$ distill 3 batches of final key separately.
The advantage in such a case is unditional, there is no loss for
whatever channel. Now we start to consider something more subtle:
the advantage conditional on the prior information of the asymmetry property
of noise of the physical channel.
\section{ 6-state protocol with 2-way classical communications.}
The symmetric channel noise for a 6-state protocol  is defined
by $q_{x0}=q_{y0}=q_{z0}$, if this condition is broken, we regard it
as an asymmetric channel for 6-state protocols.
In the
standard 6-state protocol\cite{6state,gl,chau}, symmetrization is
used, i.e., the key-bits are equally consisted by $X-,Y-,Z-$bits. When the channel noise itself is symmetric, i.e.,
$q_{x0}=q_{y0}=q_{z0}$, a 6-state protocol can have a higher noise
threshold than that of a 4-state protocol. This is because in the
6-state protocol, the $\sigma_y$ type of channel error rate is
also detected. In removing the bit flip error, $\sigma_y$ error is
also reduced, therefore the phase-flip error is partially removed. 
However, in a 4-state protocol, $\sigma_y$ error is
never tested therefore we have to assume the worst situation that
$q_{y0}=0$\cite{gl}.

We shall show that one can have a higher tolerable channel errors if
one modify the existing protocols, given the asymmetric channel(i.e.,
the channel with its $Y$-bits flipping error being different from that of $X$-bits
or Z-bits.)  For example, in the case that $q_{y0}=0$ and
$q_{x0}=q_{z0}$. The different types of error rates to the transmitted
qubits are
$$
q_{x}=q_{x0}, q_{y}=0,q_z=q_{z0}
$$ for those $Z-$bits;
$$
q_{x}=q_{z0}, q_{y}=0,q_z=q_{x0}
$$
for those $X-$bits and
$$
q_{x}=q_{z0}, q_{y}=q_{x0},q_z=0
$$ for $Y-$bits.  The average error rates over all transmitted bits are:
\begin{eqnarray}
\bar q_x=q,\bar q_y=q/3, \bar q_z=2q/3, q=q_{x0}=q_{z0}\label{trans}
.\end{eqnarray}
With such a fact, the  threshold of total channel noise 
$q_{t0}=(q_{x0}+q_{y0}+q_{z0})$ 
for the protocol \cite{chau} is $41.4\%$,
same with the case with symmetric noise\cite{chau}. Actually, by our
numerical calculation we find that the threshold of total channel noise 
for Chau protocol\cite{chau} is almost unchanged with whatever value of 
$q_{y0}$.
 However, if $all$ key bits were prepared in
$Y-$basis (the basis of $\{|y\pm\rangle=\frac{1}{\sqrt
2}(|0\rangle\pm i|1\rangle)\}$), there would be no $\sigma_z$ type
of error therefore one only needs to correct the bit-flip error.
To see this we can regard a 
$Y-$qubit as a qubit treated in the following order
\\{\it prepared in  $Z$ basis, $T$ transform, transmitted over the noisy 
channel, $T^{-1}$ transform, measurement in $Z$ basis.}
Here $T=\frac{1}{\sqrt 2}\left(\begin{array}{cc}
1 & i\\1 & -i
\end{array}\right)$, it changes states $|0,1\rangle$
into $|y\pm\rangle$. The following facts 
\begin{eqnarray}
T\sigma_x T^{-1}=\sigma_y; T\sigma_y T^{-1}=\sigma_z;
T\sigma_z T^{-1}=\sigma_x
\end{eqnarray} 
cause the consequence that
a channel flip of the type $\sigma_x,\sigma_y,\sigma_z$ will cause an
error to $Y-$bits in the type of $\sigma_y,\sigma_z,\sigma_x$ respectively.
 With
this fact, if the channel error of $p_{y0}$ is 0, the $\sigma_z$
type of error to the $Y-$qubit is also 0. Using
the iteration formula given in Ref\cite{chau}, once all bit-flip
error is removed, all errors are removed. Therefore the
error rate threshold is
\begin{eqnarray}
Q_x=Q_z=25\%,
\end{eqnarray}
i.e., a total error rate of $50\%$.  
In practice, it is not likely that $\sigma_y$ type of
channel flip is exactly 0. Numerical calculation (Fig. 1) shows that, using
$Y-$basis as the only basis for all key-bits always has an advantage provided
that the channel flipping rate satisfies 
$q_{y0}<q_{x0}=q_{z0}$. For the purpose of improving the noise
threshold, we propose the following protocol:
\begin{itemize}
\setlength{\itemsep}{-\parskip}

\item [\bf 1:] Alice creates a random binary string $b$ with
$(6+\delta)n$ bits.

\item [\bf 2:]Alice generates $(6+\delta)n$ quantum state according
to each elements of $b$. For
    each bit in $b$, if it is 0, 
she produces a quantum state  of either $|0\rangle$, 
or $|+\rangle$ or $|y+\rangle$
with probability $1/4,1/4,1/2$, respectively; if it is  $1$, 
she creates a quantum state  $|1\rangle$, or $|-\rangle$ or $|y-\rangle$
with probability $1/4,1/4,1/2$, respectively.

\item [\bf 3:] Alice sends all  qubits to Bob.

\item [\bf 4:] Bob receives the $(6+\delta)n$ qubits, measuring
each one
    in basis of either $\{|0\rangle$,$|1\rangle\}$ or $\{|\pm\rangle\}$
        or $\{|y\pm\rangle\}$  randomly, with equal probability.

\item [\bf 5:] Alice announces basis information of each qubits.

\item [\bf 6:] Bob discards all those qubits he measured in
a wrong basis.  With high probability, there are at
    least $2n$ bits left (if not, abort the protocol).  Alice
        randomly chooses  $n$ $Y$-bits to use for the distillation of
final key, and
         use the remained $n$ bits as check bits. (Among all the check bits,
         approximately
$n/3$  are $X$-bits, $n/3$ are Z-bits, and $n/3$ are $Y$-bits.)

\item [\bf 7:] Alice and Bob announce the values of their check
bits.
    If too many of them are different, they abort the protocol.

\item [\bf 8:] Alice randomly group the key bits with each group
consisting 2 bits.  Alice and Bob compare the parity values on
each side to each group. If the values agree, they discard one bit
and keep the other one. If the value disagree, they discard both
bits of that group. They repeatedly do this for a number of
 rounds until they believe they can find a certain integer $k$
 so that both bit-flip error and
 phase-flip error are less than $5\%$ after the following step is
 done.

\item [\bf 9:] They randomly group the remained key bits with each
group consisting $k$ bits. They use the parity value of each group
as the new bits after this step.

\item [\bf 10:] Alice and Bob use classical CSS code to distill
the final key.
\end{itemize}
Remark 1: Step 9 is to remove the phase flip error of the final key. 
Although in
the extreme case that the phase flip error rate is always 0 if
initially the $\sigma_y$ type of error is 0, however, in practice
the initial $\sigma_y$ type of error rate is not exactly 0. Even
in the case the tested error rate on check bits is zero, we still
have to assume a small error rate on the key bits
to increase the confidence
level.
\\Remark 2: The above protocol is unconditionally secure. This
means, under whatever type of intercept-and-resend attack, Eve's
information to the final key is exponentially small. The security
proof can be done through the purification and reduction procedure
given by Ref.\cite{shorpre}. The only thing that is a bit different
here is that Alice and Bob will take measurement in $Y$ basis to
make the final key, after the distillation.
\\Remark 3: We don't recommend to use the above protocol blindly.
Before doing the QKD, the users should test their physical channel and
decide whether there is an advantage. Numerical calculation shows that,
In the case of $q_{x0}=q_{z0},
q_{y0}<q_{x0}$, our protocol always has a higher error rate
threshold than the corresponding 6-state protocol with key bits'
bases equally distributed in all 3 bases. This fact is shown in Fig.(1).
\\Remark 4: The conditional advantage require the users first test
the properties of the physical channel before doing QKD. And we assume
that the physical channel itself is stable. Note that we don't 
require anything
for Eve's channel. As we have discussed in the beginning, physical
channel is in general different from Eve's channel since Eve
may  take over the whole channel only at the time Alice and Bob do
the QKD. However, if Eve wants to hide her presence, she must
respect the expected results of the error test in the protocol. 
In our protocol, Eve's operation must not change the error rates
of the physical channel, though she can change the error pattern. 
Since if these values
are changed, Alice and Bob will find that their error test result is
much different from the expected one therefore Eve cannot hide here presence.
\\One may still worry whether the conditional advantage in our protocol
is really useful, especially for the conditional threshold part.
To make it clear, we consider a specific game. Suppose now Alice and Bob
are prisoned in two separate places. They are offered a chance to be freed
immediately. The rule is set as this: If they can make an unconditionally
secure final key, they will be freed immediately. If any third party
obtained a non-negligible amount of information of the final key, 
they will be shot
immediately. Suppose both Alice and Bob want to be freed immediately,
but they take the value that being alive is more important than 
freedom\cite{free}. 
The noise of the physical is known: the $p_y=0; p_x=p_z=22\% $.
In such a case, those conditionally secure protocol with untestable conditions
cannot be used, since those protocols will bring the risk to
Alice and Bob of being shot. 
For example, protocol $T$ is conditionally secure with
individual attack, but we don't know how to see whether Eve has only used
the individual attack. Even though $T$ has a very large tolerable 
channel noise, Alice and Bob cannot use this protocol because
they have a risk to be shot immediately.
Previously known unconditionally secure protocols will not bring 
the risk of being
shot to Alice and Bob, but those protocols cannot bring liberty to 
Alice and Bob, since none of them can tolerate 
such a high channel flipping error.
However, the protocol in this work can help Alice and Bob to be freed
without any non-negligible risk of being shot. In such a case, our protocol is
the only protocol that may help Alice and Bob while all previously known
unconditionally secure protocols cannot. 

Besides the advantage of a higher tolerable error rate, there are also
advantages in the key rate of our protocol with asymmetric channel noise.
Obviously, when the error rate is higher than other protocols' threshold while
lower than our protocol's threshold, our protocol always has an advantage
in key rate. More interestingly, even in the case that the error rate is
significantly lower than the threshold of Shor-Preskill's protocol, we may
modify our protocol and the advantage in key rate may still holds. 
We modify our protocol in such a way:  take one round bit-error-rejection
with two way communication and then use CSS code to distill the final key.
As it was shown in Ref.\cite{chau}, the various flipping rates will change
by the following formulas after the bit-flip-error rejection
\begin{equation}
  \left\{ \begin{array}{rcl} 
q_I & = & \displaystyle\frac{p_{I0}^2 +
   p_{z0}^2}{(q_{I0} + q_{z0})^2 + (q_{x0} + q_{y0})^2} , \\ \\
   q_x & = & \displaystyle\frac{q_{x0}^2 + q_{y0}^2}{(q_{I0} + q_{z0})^2 +
   (p_{x0} + p_{y0})^2} , \\ \\
   q_y & = & \displaystyle\frac{2q_{x0} q_{y0}}{(q_{I0} + q_{z0})^2 + (q_{x0} +
   q_{y0})^2} , \\ \\
   q_z & = & \displaystyle\frac{2q_{I0} q_{z0}}{(q_{I0} + q_{z0})^2 + (q_{x0} +
   p_{y0})^2} .
  \end{array}
  \right. \label{errorrate}
 \end{equation}
Also, it can be shown that, the number of remained pairs is
\begin{eqnarray}
f=\frac{1}{2}\frac{1}{(q_{I0} + q_{z0})^2 + (q_{x0} +
   p_{y0})^2}.
\end{eqnarray}
The key rate of our protocol is given by
\begin{eqnarray}
R=f\cdot (1+q_x \log_2 q_x + q_y \log_2 q_y + q_z \log_2 q_z + q_I \log_2 q_I)
\end{eqnarray}
We shall compare this with the key rate of our modified six-state 
protocol where key bits are equally distributed over 3 different bases
but we distill 3 batches of final key,  i.e. 
\begin{eqnarray}
r'=1+ q_{x0} \log_2  q_{x0} +  q_{y0} \log_2  q_{y0} 
+  q_{z0} \log_2 q_{z0} + q_{I0} \log_2 q_{I0}.
\end{eqnarray}
We need not compare our results with the standard 6-state protocol\cite{6state}since its key rate given by eq.(\ref{6mix}) is superemed by eq.(\ref{6sep}).
The numerical results are given in Fig.(2).
\begin{figure}
\epsffile{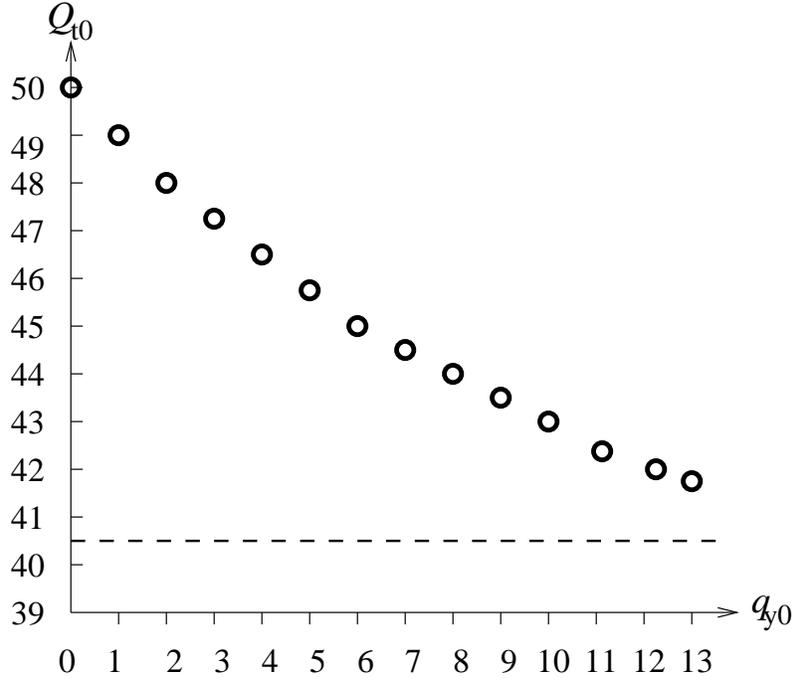}
\vskip 15pt
\caption{ Comparison of channel error threshold of different protocols.
All values are in the unit of one percent. $Q_{t0}$ is the threshold value
of total channel noise given certain value of $q_{y0}$. In calculating 
$Q_{t0}$, we assume $q_{x0}=q_{z0}$.
The dashed line is the threshold value
of Chau protocol, the circled curve is the threshold of the six state protocol
given in this work.}
\end{figure} 
\begin{figure}
\epsffile{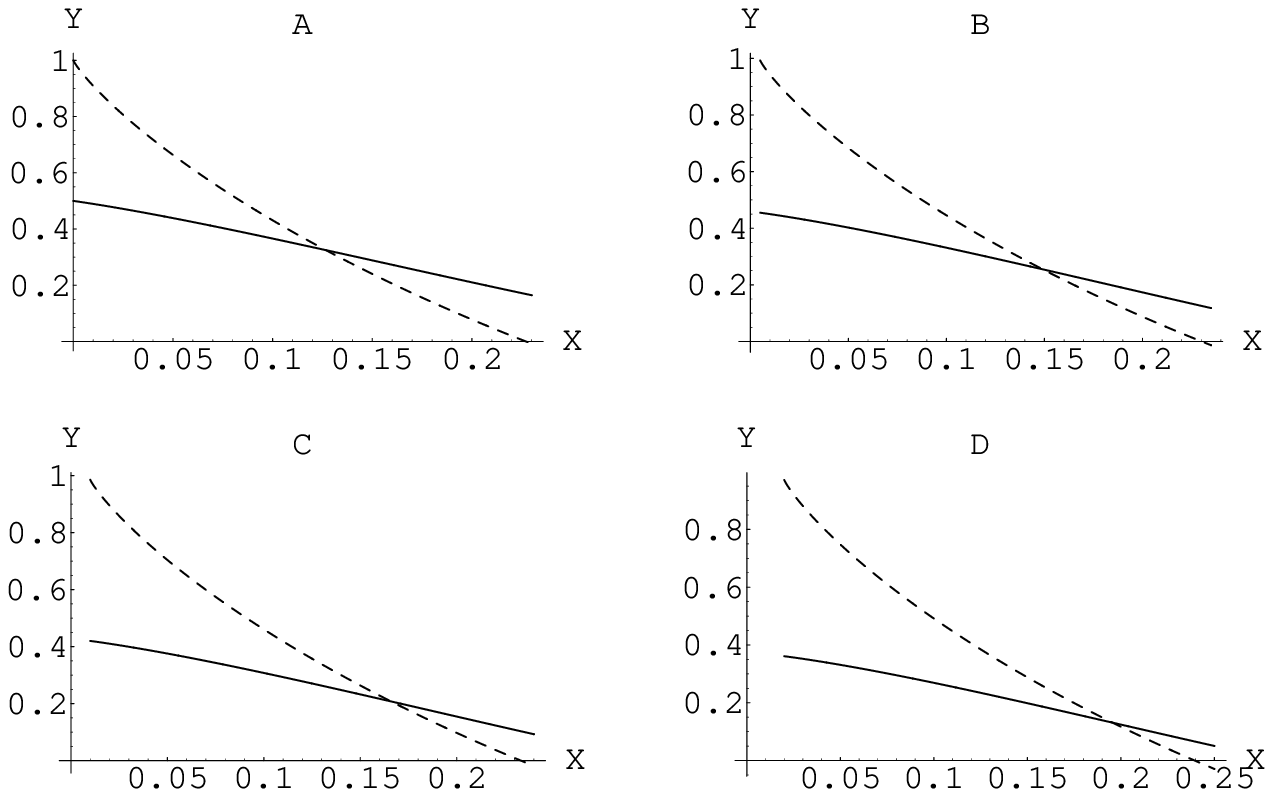}
\vskip 15pt
\caption{ Comparison of the key rate of our protocol( the solid line) 
and six-state Shor-Preskill protocol (the dashed line). The Y-axis is for the key rate, X axis is $q_{x0}+q_{y0}+q_{z0}$; A, B, C, D are for the cases of
$q_{y0}=0, 0.5\%,1\%,2\%$, respectively. In all cases we have assumed 
$q_{x0}=q_{y0}$. We can see that, if $q_{y0}$ is small,
our protocol can have a higher efficiency than six-state Shor-Preskill protocol even the total error rate is significantly lower than the threshold value
of the six-state Shor-Preskill protocol. }
\end{figure} 
\section{ Summary} In summary, We have shown that, given the asymmetric
channel flip rate one can have advantages in tolerable flip rates and
efficiency, if one uses a single basis for the key-bits.
We have demonstrated this point by both 4-state protocol with CSS-code and
the 6-state protocol with 2-way communication.
It should be interesting to investigate the most general case
that $p_{x0},p_{y0},p_{z0}$ are all different for the case of 6-state protocol.
\\{\bf Acknowledgement:} I thanks Prof Imai H for support. 
 Valuable conservations
with H. F. Chau is also gratefully acknowledged.


\begin{thebibliography}{99}
\bibitem{free}  We don't insist
on the value that being alive is more important than freedom for humen beings in general, though we assume such a value for Alice and Bob in our story here.
\end{thebibliography}
\end{document}